\documentclass[useAMS,usenatbib]{mn2e}
\usepackage{graphicx}
\usepackage{amssymb}
\usepackage{subfigure}
\usepackage{MnSymbol}

\newcommand{\idest}{i.e.}           
\newcommand{\Msun}{\mbox{\,$M_{\odot}$\/}}          
\newcommand{\oversim}[2]{\lower0.5ex\vbox{\baselineskip=0pt\lineskip=0.2ex
     \ialign{$\mathsurround=0pt #1\hfil##\hfil$\crcr#2\crcr\sim\crcr}}}

\usepackage{graphicx}
\usepackage{nicefrac}
\usepackage{lscape}
\usepackage{amsmath}

\title[Testing the universality of star formation II]{Testing the universality of star formation - II. 
\\ Comparing separation distributions of nearby star-forming regions and the field}
\author[R.~R.~King, S.~P.~Goodwin, R.~J.~Parker \& J.~Patience]{Robert R.~King$^{1}$\thanks{E-mail:
    rob@astro.ex.ac.uk}, Simon P.~Goodwin$^2$, Richard J.~Parker$^{3}$ and Jenny Patience$^{1,4}$\\
$^{1}$Astrophysics Group, College of Engineering, Mathematics and Physical Sciences, University of Exeter, Stocker Road, Exeter EX4 4QL, UK\\
$^{2}$Department of Physics and Astronomy, University of Sheffield, Hicks Building, Hounsfield Road, Sheffield, S3 7RH, UK\\
$^{3}$Institute for Astronomy, ETH Z\"{u}rich, Wolfgang-Pauli-Strasse 27, 8093 Z\"{u}rich, Switzerland\\
$^{4}$School of Earth and Space Exploration, ASU, PO Box 871404, Tempe, Arizona, USA}

\begin{document}

\date{Accepted 2012 September 11. Received 2012 September 11; in original form 2012 August 8}

\pagerange{\pageref{firstpage}--\pageref{lastpage}} \pubyear{2012}

\maketitle

\label{firstpage}

\begin{abstract} 

We have measured the multiplicity fractions and separation distributions of seven young star-forming
regions using a uniform sample of young binaries. Both the multiplicity fractions and separation
distributions are similar in the different regions. A tentative decline in the multiplicity fraction
with increasing stellar density is apparent, even for binary systems with separations too close
(19--100\,au) to have been dynamically processed. The separation distributions in the different
regions are statistically indistinguishable over most separation ranges, and the regions with higher
densities do not exhibit a lower proportion of wide (300--620\,au) relative to close (62--300\,au)
binaries as might be expected from the preferential destruction of wider pairs. Only the closest
(19--100\,au) separation range, which would be unaffected by dynamical processing, shows a possible
difference in separation distributions between different regions.  The combined set of young
binaries, however, shows a distinct difference when compared to field binaries, with a significant
excess of close (19--100\,au) systems among the younger binaries.
Based on both the similarities and differences between
individual regions, and between all seven young regions and the field, especially over separation
ranges too close to be modified by dynamical processing, we conclude that multiple star formation is
not universal and, by extension, the star formation process is not universal.

\end{abstract}


\section{Introduction}

Observations of the field show that at least one-third, and possibly over half of stars are in
binary systems\footnote{There is evidence that a large number of systems are higher-order multiples
rather than simply binaries \citep{Law:2010}.  Although the majority of the observations in this
paper are binary systems, the reader should bear in mind that this may be too simplistic a view.}
\citep[e.g.][]{Michell:1767,Heintz:1969,Abt:1976,Duquennoy:1991,
Fischer:1992,Lada:2006,Bergfors:2010,Raghavan:2010}. Young stars appear to have an even greater
multiplicity than field stars, with some star-forming regions having almost all their stars in
multiple systems \citep[e.g.,][]{Mathieu:1994,Patience:2002}.  This suggests that binaries are a
very significant, if not the most significant, mode of star formation.  Any theory of star formation
must explain the properties of the binary systems observed, such as the multiplicity fraction and
separation distribution, and why the degree of multiplicity apparently falls between young star
forming regions and the field.

An important question in star formation is the universality of the star formation process: do stars
(often multiples) always form in fundamentally the same way?  Are massive star-forming regions such
as 30 Doradus, which form $>10^5 M_\odot$ of stars, in some way just scaled-up versions of sparse,
low-density regions such as Taurus, or are they fundamentally different?  Are regions with similar
densities and masses, such as Ophiuchus and IC~348, basically the same?  That the IMF appears to be
universal \citep{Bastian:2010} might argue for universal star formation, but examining binary
properties in different regions provides another method to test this hypothesis.

Unlike the apparently universal IMF, the binary properties of the field and young
star-forming regions appear to be different. The most common interpretation of this
interesting discrepancy is that young binaries are `processed' -- that is, some proportion
of young multiple systems are destroyed by dynamical interactions, and/or that the decay of
higher-order multiples dilutes the multiplicity of stars over time
\citep{Heggie:1975,Hills:1975,Kroupa:1995b,Kroupa:1995a,Goodwin:2005,Parker:2009,Kroupa:2011,Marks:2011a}.

Some degree of processing {\em must} occur, but its importance is an open question.
Processing occurs more efficiently in denser environments: there are more, and closer, encounters
between systems in denser environments.  Therefore if most stars are born in (or go through) a dense
environment, then processing may be very important \citep[e.g.][]{Parker:2009,Marks:2012}.

This is the second in a series of papers in which we analyse and interpret observations of binary
systems in the context of examining the universality (or otherwise) of star formation. A serious
complication we face in analysing observations of binary systems is that different surveys of
different populations have different selection effects.  Surveys can be sensitive to different
separation ranges, primary masses, and minimum companion masses.  To alleviate this problem, we
presented datasets for 5 young star-forming regions with which we constructed uniform samples to allow
direct and meaningful comparisons of the multiplicity fractions \citep[][hereafter
Paper I]{King:2012}.  This was done for five regions: Taurus, Chamaeleon~I, Ophiuchus, IC~348, and the
Orion Nebula Cluster (ONC).  In this paper we add two more regions -- Corona Australis (CrA) and Upper Scorpius (USco).

In this paper, we examine the binary fractions and separation distributions of binary systems in our
different regions.  The processing of binaries is expected to lower the binary fraction, and this
has been the focus of much previous work.  However, processing should also alter the separation
distribution of binaries: wider binaries should be more susceptible to destruction, possibly leading
to fewer wider binaries in denser regions than initially formed in those environments.

\section{Binary properties and processing}
\label{processing}

In this section we will briefly review how binary properties are characterised and how binary
populations are expected to be processed.

\subsection{Binary properties}

The binary properties of a particular population can be characterised by a number of quantities. The
multiplicity, or binary fraction, is a measure of the fraction of stars in binary (or higher-order)
systems and can be formulated in a number of useful ways \citep[see][]{Reipurth:1993}. In this paper
we will use the multiplicity fraction (MF) defined as
\begin{equation}
{\rm MF} = \frac{{\rm B} + {\rm T} + {\rm Q}}{{\rm S} + {\rm B} + {\rm T} + {\rm Q}}
\end{equation}
where S, B, T and Q are the numbers of single, binary, triple and quadruple systems, respectively.

For any given system, the four most important properties are the primary mass, $M_p$, and the
secondary mass, $M_s$, which give the mass ratio $q = M_s/M_p$, and the two orbital elements of
semi-major axis, $a$, and eccentricity, $e$.  Within any population, there will be a distribution of
$M_p$ and $M_s$ (related to the IMF), and of both $a$ and $e$.

It is important to remember that for visual binaries (especially those considered in this
paper\footnote{Although the observational samples used include optically- and speckle
interferometry-resolved binaries, we refer to them collectively as visual binaires.}) we observe the
instantaneous projected separation on the sky, $s$.  This depends on $a$ and $e$, but also on the
unknown phase, orientation, and inclination of the orbit.  In this paper, we analyse the {\em
instantaneous projected separation distribution}, that is, the distribution of separations on the
sky.  This is the raw observed quantity and will be related to $a$ and $e$ in a non-trivial way
\citep[see][]{Maxted:2005,Allen:2007}.

\subsection{Binary processing}

Multiple systems can be processed in two ways: externally and internally. Internal processing is
the rapid decay of unstable high-order multiples.  Many systems with $N>2$ will be unstable and
decay on a timescale of roughly 100 crossing times \cite[see][]{Anosova:1989, Sterzik:1998}. 
Usually the lowest-mass member is ejected and the remaining system has lower energy and so becomes
closer (harder, see below). This will change the binary fraction of a region (e.g. a triple becomes
a binary and a single) and alter the separation distribution \citep[see][]{Goodwin:2005}.  However,
this process is extremely rapid (timescales of $<10^5$\,yrs) and will occur during the early stages
of star formation \citep{Goodwin:2005}, probably at much younger ages than the young stars in our
sample.

External processing from encounters with other stars/systems is more interesting as it should occur
differently depending on the environment.  Binaries can be divided into two broad categories: `hard'
and `soft' \citep{Hills:1975,Heggie:1975}.  Hard binaries are so strongly bound that it is highly
unlikely that an encounter with enough energy to destroy them will occur.  Soft binaries are so
weakly bound that they are almost certain to be destroyed.  In between soft and hard binaries are
`intermediate' binaries whose destruction depends on chance as to whether they have a destructive
encounter or not \citep[see][]{Parker:2012}.

The chance of a binary surviving in an environment depends on the binding energy of the binary and
the frequency and energy of encounters (to first order set by the density and age of the
environment). Roughly speaking, an encounter is destructive if the relative velocity of the
encounter is greater than the orbital velocity of the binary.

The boundary between hard and soft binaries is the separation significantly below which dynamical
destruction is  very unlikely (hard), and significantly above which dynamical destruction is almost
certain (soft).  For a $0.5$--$0.5 M_\odot$ binary that encounters a $0.5 M_\odot$ star, the
hard-soft boundary, $a_{\rm hs}$, is approximately
\begin{equation}
a_{\rm hs} \sim 450 \left( \frac{\sigma}{{\rm km~s}^{-1}} \right)^{-2}
\,\,\,\,{\rm au},
\end{equation}
where $\sigma$ is the typical encounter velocity \citep[see][]{BandT}. The hard-soft boundary will
be at closer separations if the binary is more massive, or if the encounter is with a higher-mass
star.

The timescale, $t_{\rm enc}$, for an encounter at a distance, $d$, in an environment with number
density of systems/stars, $n$, with velocity dispersion, $\sigma$, is
\begin{equation}
t_{\rm enc} \sim 10^{10} \left( \frac{n}{{\rm pc}^{-3}} \right)^{-1}
\left( \frac{\sigma}{{\rm km~s}^{-1}} \right)^{-1} \left(
\frac{d}{1000\,{\rm au}} \right)^{-2} \,\,\,\,\, {\rm yrs}.
\end{equation}
For a typical young star-forming region with $\sigma \sim 2$ km s$^{-1}$ and $n = 10^2$\,pc$^{-3}$,
we expect an encounter at 1000\,au every $\sim 50$\,Myr.  Alternatively, one in every 50
systems/stars will have an encounter at 1000\,au every Myr.

In summary, we expect dynamical processing to depend on the environment, with more numerous and more
destructive encounters in denser, higher velocity dispersion environments.  Hard (close) binaries
will almost never be destroyed, but the definition of hard depends on the density of the
environment. It should be noted that the destruction of binaries around the hard-soft boundary is a
stochastic process depending on the exact history of a particular binary.  However, we expect denser
environments to be more destructive and to preferentially destroy wider binaries
\citep[see][]{Kroupa:1995a,Marks:2011a,Marks:2011b,Parker:2012}. Therefore, if binary formation is
universal, then we would expect to see both a lower binary fraction, and fewer wider binaries in
dense regions than in low-density regions \citep[see][]{Marks:2011a}. However, it should also be
noted that it is the {\em maximum} density a region reaches/reached that sets the degree of
destructiveness of encounters -- not necessarily the current density
\citep[see][]{Parker:2009,Parker:2011c} -- and therefore differences/similarities between regions may be
due to {\em past} differences/similarities.

It is interesting to examine a binary separation range which should be unaffected by dynamical
processing.  The velocity dispersion in our regions are at most $\sim 2$\,km~s$^{-1}$
\citep{Frink:1997,Kraus:2008b} which suggests that binaries with separations of $< 100$\,au should
be hard in all of our clusters.  This means we should be observing binaries almost certain to 
be unaffected by dynamical processing and we will refer to these binaries as `pristine'.  For five
of our regions (Cha, CrA, Oph, Tau, and USco) we have data for the range 19--100\,au.


\begin{table}
  \centering
  \caption{A summary of the separation ranges and contrasts from
  each binary survey used.}
\label{tab:contrasts}
\begin{tabular}{llllr}
  \hline
  \hline
  Region 	&  \multicolumn{2}{c}{Separation Range} &  Contrast & Reference  	\\
  
  		& (arcsec)	& (au)		&  &\\
  \hline
  Taurus 	&  0.13--13.	&  18--1820	& $\Delta K \le2.5$			& 1	 \\
  Oph       &  0.13--6.4	&  17--830  & $\Delta K \le2.5$			& 2  \\
  Cha~I	   &  0.10--6.0	&  16--960	& $\Delta K \le3.1$			& 3  \\
  CrA 		&  0.13--6.0	&  17--774  & $\Delta K \le2.5$			& 4  \\
  USco 		&  0.13--6.0	&  19--870  & $\Delta K \le2.5$			& 5  \\
  IC~348 	&  0.10--8.0	&  32--2530 & $\Delta H \le6.5$			& 6  \\
  ONC 		&  0.15--1.5	&  62--620	& $\Delta H \alpha \le5.0$	& 7  \\
  \hline
\end{tabular}
{\it References}: (1) \citet{Leinert:1993}, (2) \citet{Ratzka:2005}, (3) \citet{Lafreniere:2008},
(4) \citet{Kohler:2008}, (5) \citet{Kohler:2000}, (6) \citet{Duchene:1999}, (7) \citet{Reipurth:2007}  
\end{table}

\section{Binary star samples}
\label{sect2}


In Paper I, we compared the binary surveys of young stars in five well-studied regions: Cha~I
\citep{Lafreniere:2008}, Ophiuchus \citep{Ratzka:2005}, Taurus \citep{Leinert:1993}, IC~348
\citep{Duchene:1999} and the ONC \citep{Reipurth:2007}. These regions were chosen both because they
had been surveyed for binary companions and due to their well known stellar membership, which was
required to determine their stellar densities.  In this paper we have also included binary surveys
of Upper Scorpius \citep[USco,][]{Kohler:2000} and Corona Australis \citep[CrA,][]{Kohler:2008}. 
Unfortunately, the membership of these two new clusters is poorly known and so determining their
densities is very difficult.

\subsection{Binary surveys used}

The binary surveys of Cha~I \citep{Lafreniere:2008}, IC~348 \citep{Duchene:1999} and the ONC
\citep{Reipurth:2007} were discussed in detail in Paper I and the same data are used in this study.
However, the sample of stars considered in Taurus and Ophiuchus differs here from that presented in
Paper I. For Taurus, we previously focused on the area of the northern filament, but here we have
used the full survey of 104 targets from \citet{Leinert:1993}. For Ophiuchus, previously we had
considered only those stars which were listed as members by \citet{Wilking:2008}. Here, we make use
of the full binary survey of members identified by \citet{Ratzka:2005}.

\subsubsection{Corona Australis}
\label{sect:cra}

The Corona Australis (CrA) star-forming region has been recognised as a host to young stars since at
least the 1960s \citep{Herbig:1960}, but the high level of extinction in the central cloud
($A_{V}\sim 45$) has hindered the identification of the full stellar population. In their review of
this region, \citet{Neuhauser:2008} presented a compilation of the known members. Although too
little is known of the full population to determine the density of this cluster, \citet{Kohler:2008}
surveyed the binary properties of the optically visible stars with near-IR speckle interferometry
and AO imaging. From 47 target systems, they found 19 binaries and no higher-order systems. They
reported a multiplicity fraction of 36$\pm$9\,per cent within a separation range of
0.13--6.0$\arcsec$, corresponding to 17--774\,au at a distance of 130\,pc \citep{Casey:1998}, where
their observations were fully sensitive to flux ratios $\ge$0.1.  From the density of stars in the
field, none of their observed binaries are expected to be a result of projection effects. Including
previously reported single and multiple systems, \citet{Kohler:2008} determined a multiplicity
fraction of 46$\pm$10\,per cent (CSF$=52\pm10$\,per cent).

\subsubsection{Upper Scorpius}
\label{usco}

Upper Scorpius (USco) is the most compact of the three regions which constitute the
Scorpius-Centaurus (Sco-Cen) OB association, but still covers $\sim$150 deg$^2$ on the sky at a mean
distance of 145$\pm$2\,pc \citep{deZeeuw:1999}. Due to the dispersion of this cluster, searches for
members are incomplete, but $>$100 high-mass members have been identified through their proper
motion and parallaxes \citep{deZeeuw:1999} and $\sim$100 low-mass members with the {\it Einstein}
and {\it ROSAT} X-ray observations of \citet{Walter:1994} and \citet{Kunkel:1999}.
\citet{deBruijne:1999} showed that the line-of-sight depth of USco could be up to 50\,pc, adding to
the uncertainty of measures of binary separations. Until recently, USco was thought to be
approximately 5\,Myr old, considerably younger than the other subgroups of the association
\citep{deGeus:1989,Preibisch:2002}. Recently however, \citet{Pecaut:2012} revised the age to
$\sim$11\,Myr based on the luminosities of its F-type stars.

\citet{Kohler:2000} presented a binary survey of 118 young stellar systems in the Sco-Cen OB
association where, after correction for X-ray selection bias and projected companions, they found a
multiplicity fraction of 32.6$\pm$6.1\,per cent within a separation range of 0.13--6.0$\arcsec$
(corresponding to 19--870\,au at 130\,pc) and flux ratios $\ge$0.1.  For this analysis, we have
restricted the \citeauthor{Kohler:2000} sample to those 70 targets labelled as USco-A, the majority
of which were identified as members of USco by \citet{Preibisch:1999}.

\begin{figure}
    \resizebox{\hsize}{!}{\includegraphics{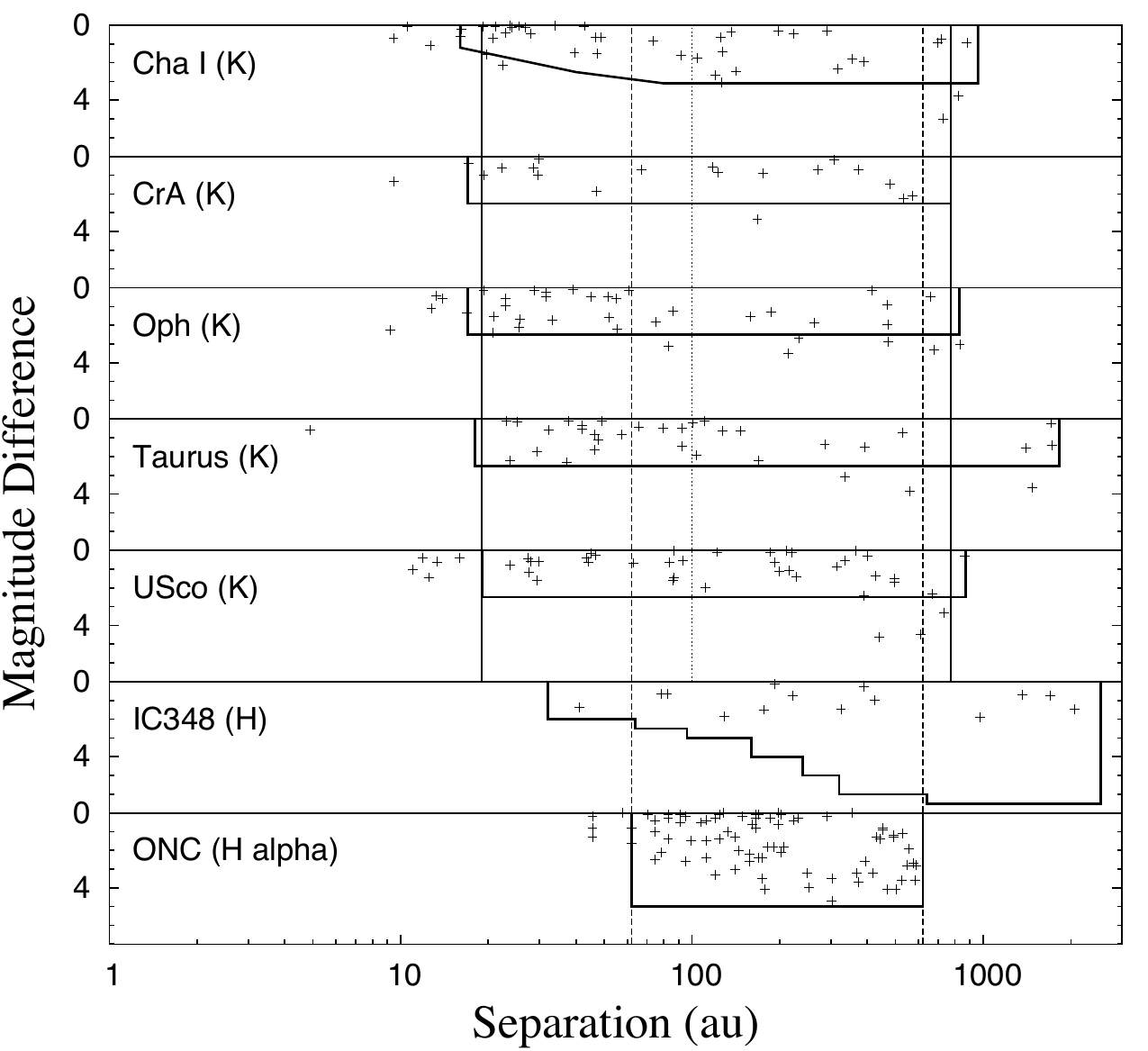}}
    \caption{The contrast of each multiple system found in the seven surveys shown as a function of
    separation. The boxes demarcate the completeness of each survey and the vertical lines mark the
    bounds of our three separation ranges, 19-774\,au (solid line), 19-100\,au (first solid and
    dotted line) and 62-620\,au (dashed line). The labels identify the clusters and the filter used
    in the observations.}
    \label{fig:contrasts}
\end{figure}

\subsection{Matching sample sensitivities}

To allow fair comparisons between the samples of binary stars from different surveys, we applied
identical cuts to the spectral type of the target stars, the contrast between primary and companion, and
the projected physical separation of the binary.  These limits were chosen such that all surveys
were equally sensitive to binary systems, allowing us to compare what has been observed without
having to apply uncertain corrections.

As in Paper I, we consider only targets with spectral types from G5 to M6 (limiting the primary
masses to $\sim$0.1--3.0$\Msun$) and binaries with a contrast of $\Delta K \le 2.5$ (or
equivalently, $\Delta H \le 2.7$, $\Delta H \alpha \le 5.0$). To maximise the separation range
probed, we present a comparison over three different separation ranges. The widest separation range
over which this contrast is achievable is 19--774\,au, applicable to all regions except IC~348 and
the ONC (the two most distant regions). As described in Sect.\,\ref{processing}, we compare five
regions over the `pristine' range of 19--100\,au, unaffected by dynamical evolution, and for all
seven regions, we also compare companion separations over the common separation range of
62--620\,au.

\subsection{Spectral type distributions}
\label{sect:Spectral type distributions}

Observations of binarity across the full range of stellar and sub-stellar masses suggests that the
binary fraction increases with stellar mass and the peak of the separation distribution moves
outward
\citep{Burgasser:2006,Close:2003,Basri:2006,Fischer:1992,Duquennoy:1991,Preibisch:1999,Mason:1998}.
This implies that the separation distribution in a population is sensitive to the masses of the
stars surveyed. To address this concern, we have investigated the distribution of spectral types
among the target stars in all seven binary surveys.  To build comparable samples, we have
restricted the range of spectral types considered to G5--M6 ($\sim$0.1-3.0$\Msun$ at $\sim$1--2
Myr), except for the ONC where the targets of \citet{Reipurth:2007} are not identified.

\begin{figure}
    \resizebox{\hsize}{!}{\includegraphics{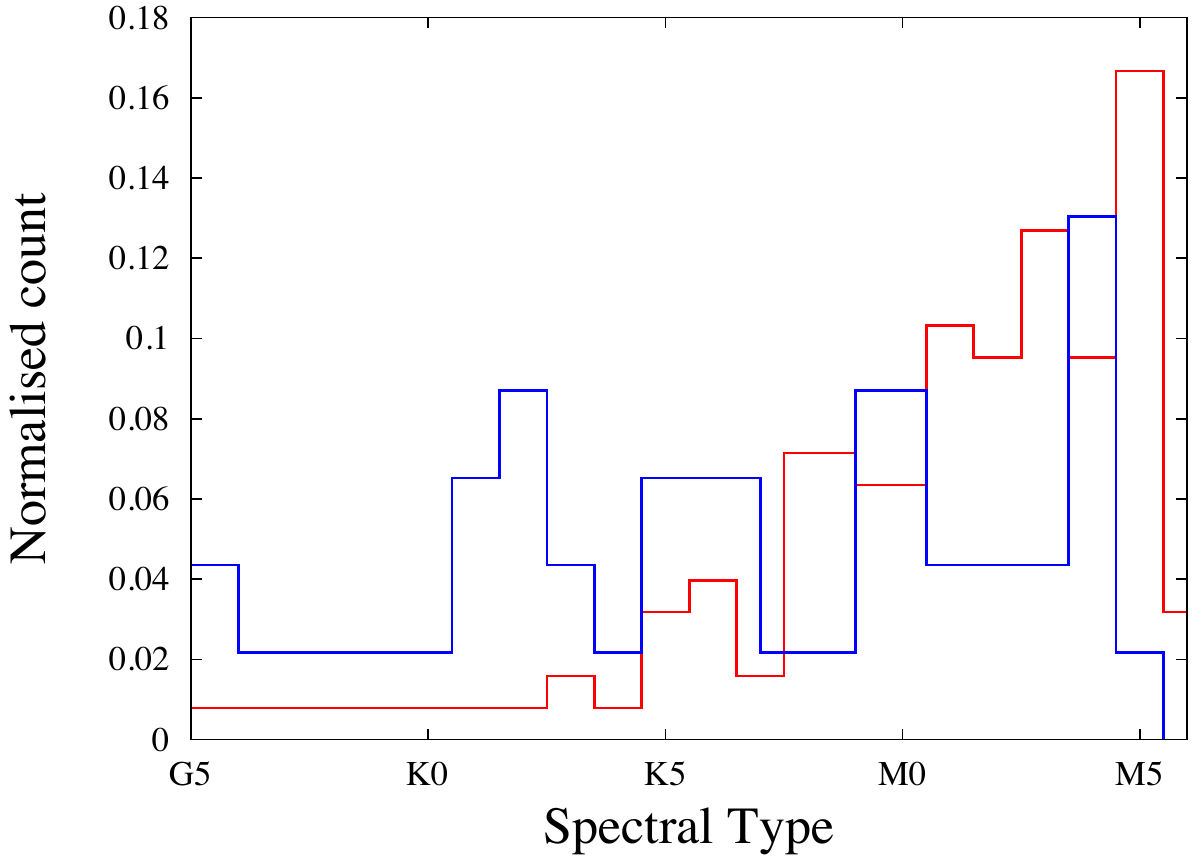}}
    \caption{The distribution of spectral types among the targets of the binary surveys of
    Cha I (red) and CrA (blue).}
    \label{fig:spt_distribution}
\end{figure}

Figure\,\ref{fig:spt_distribution} illustrates the distribution of spectral types for the targets of
the binary surveys of Cha~I and CrA (the most different distributions). Among the 6 regions with
member lists with measured spectral types, the fraction of targets with spectral types from G5 to K5
range from $\sim$10--40\,per cent.

Since all the target regions have ages of $\sim$1-2\,Myr, the targets have not yet reached the main
sequence and so evolve to earlier spectral types as they age. Using the evolutionary models of
\citet{Siess:2000} and the colour-effective temperature relation of \citet{Kenyon:1995}, a 2\,Myr
old G5 (K4, K5, M6) star will evolve to a spectral type of A0 (F6, G6, M6) by 1\,Gyr. The majority
of the $\sim$2\,Myr old cluster primaries (G5--M6) therefore cover approximately the same range in
mass as the field binary surveys of \citet{Duquennoy:1991} and \citet{Fischer:1992}.

\subsection{Cumulative distribution functions}

\begin{figure}
    \resizebox{\hsize}{!}{\includegraphics{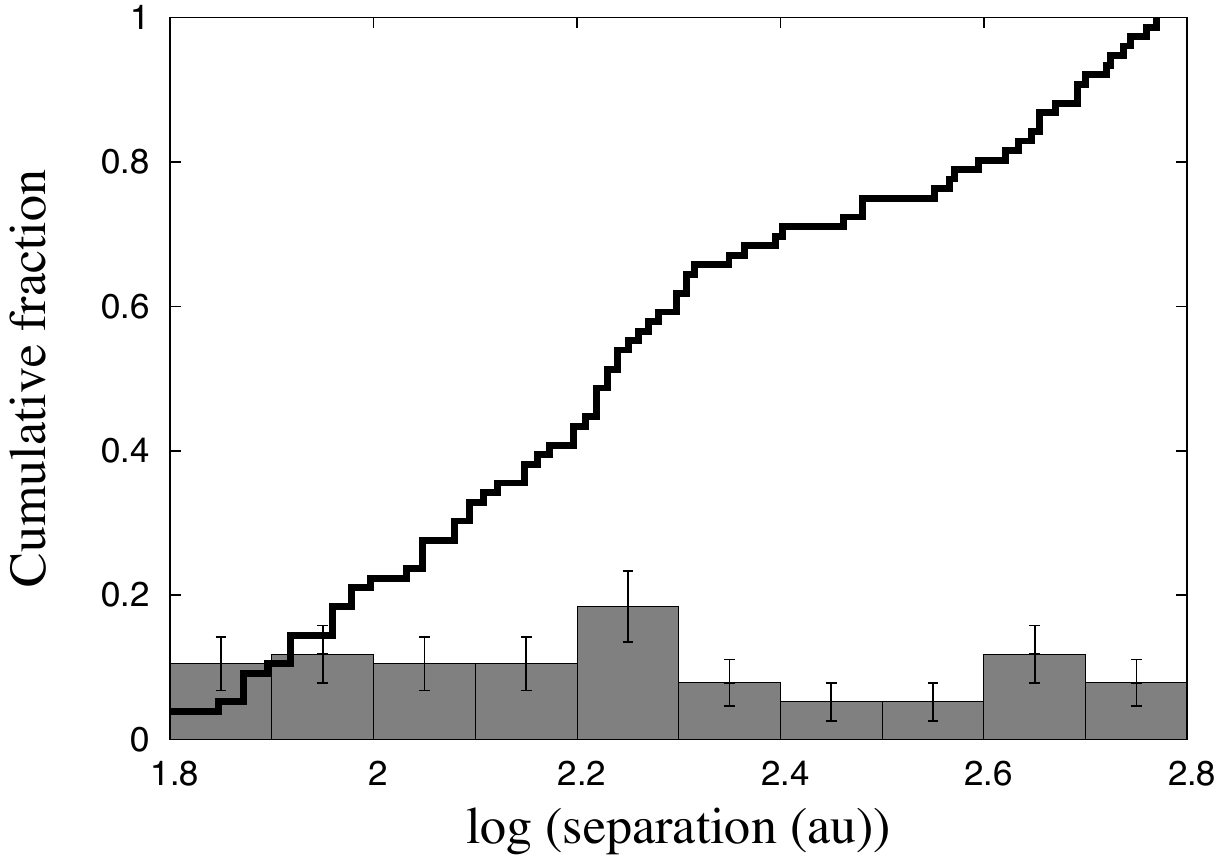}}
    \caption{A comparison of the cumulative separation distribution and a binned histogram 
    for the binary companions in the ONC within the separation range 62-620au.}
    \label{fig:comp_sep_distribution}
\end{figure}

In this paper, we concentrate on the separation distributions of multiple systems.  For visual
binaries, observations record the instantaneous projected separation on the sky of the
system\footnote{Some nearby visual binaries have been observed sufficiently long (often $> 100$\,yrs)
for orbital solutions to have been found, but our point holds for the vast majority of visual
binaries, especially those in star-forming regions.}.  This separation depends not only on the
semi-major axis and eccentricity of the system, but on the unknown phase, orientation, and
inclination of the orbit. Distributions of instantaneous separations can be statistically analysed
to find the most probable underlying distributions of semi-major axis and eccentricity
\citep[e.g.][]{Maxted:2005,Allen:2007}, but in this paper we analyse the observed instantaneous
projected separations.

The distributions of binary separations are generally presented as binned histograms in
log-separation or log-period (note that determining the period relies on knowing the semi-major axis
which is not directly observed). Such histograms have the advantage of including both the binary
fraction and separation distribution in a single figure, but they also confuse the comparison
between different separation distributions.  Therefore in this paper, we will examine both the
multiplicity fraction as a function of binned separation and the cumulative distribution functions
(CDFs) of binary separations. A comparison of these approaches is shown in
Fig.\,\ref{fig:comp_sep_distribution}, which includes the separation distribution of the binary
systems in the ONC as both a binned histogram and a CDF.  Both show a relatively flat separation
distribution, but the slight excess of binaries around a separation of $\sim$200\,au (2.3 in
log$_{10}$($s$)) is more distinct in the CDF than the histogram (indeed, without the CDF this
feature is unclear).

In this paper, we analyse the separations of all {\em companions}, not simply all binary companions.
This means that multiple systems will contribute more than one separation to the CDF. We feel this
is a more consistent approach than the alternative of including only one companion per system, or
limiting any comparison to binary systems. However, we feel it is important for the future to
develop some more consistent way of including higher-order multiples, but this is beyond our remit
in this paper.


\section{Analysis}

In this section, we examine the variation of the multiplicity fraction in each of three separation
ranges of 19--774\,au, 62--620\,au, and 19--100\,au.  We then examine the binary separation
distributions in the same three separation ranges comparing young regions with each other and with
the field.

\subsection{Multiplicity fraction and density}

\begin{figure}
    \resizebox{\hsize}{!}{\includegraphics{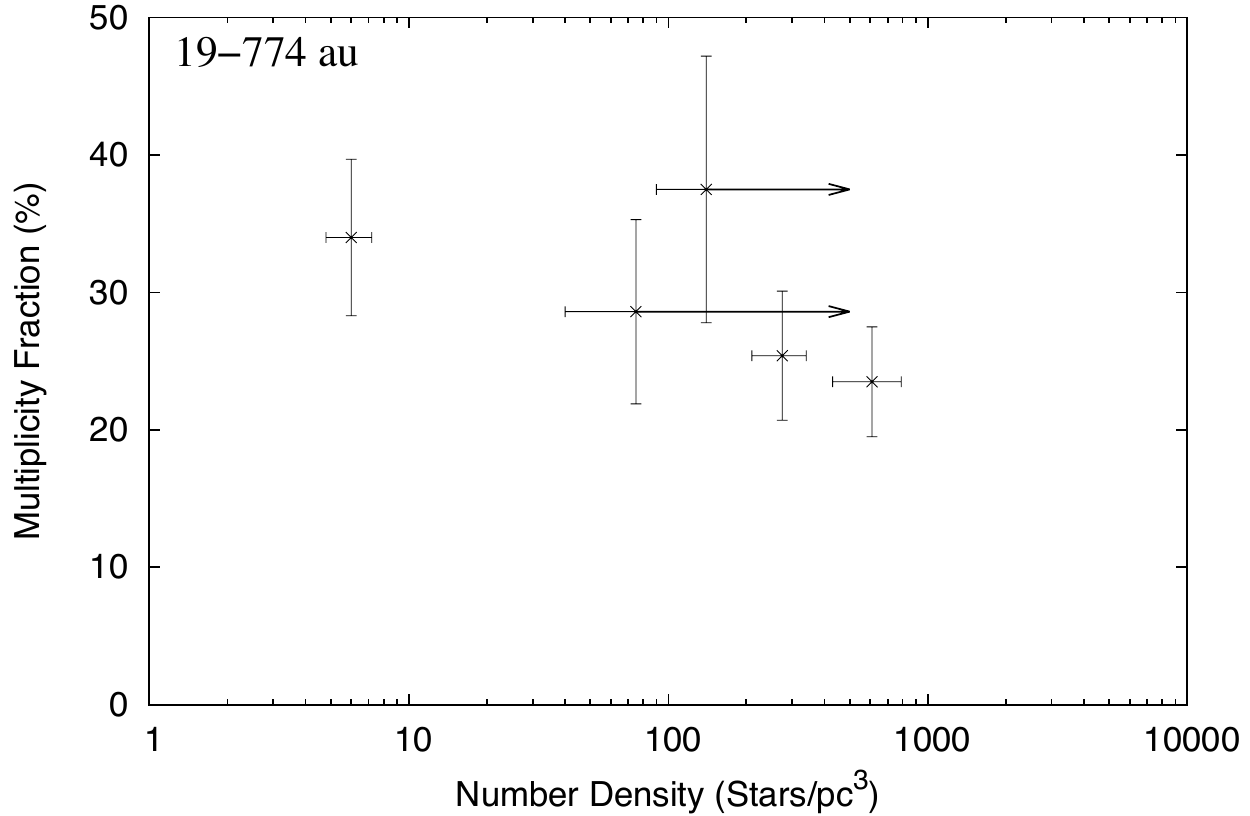}}
    \resizebox{\hsize}{!}{\includegraphics{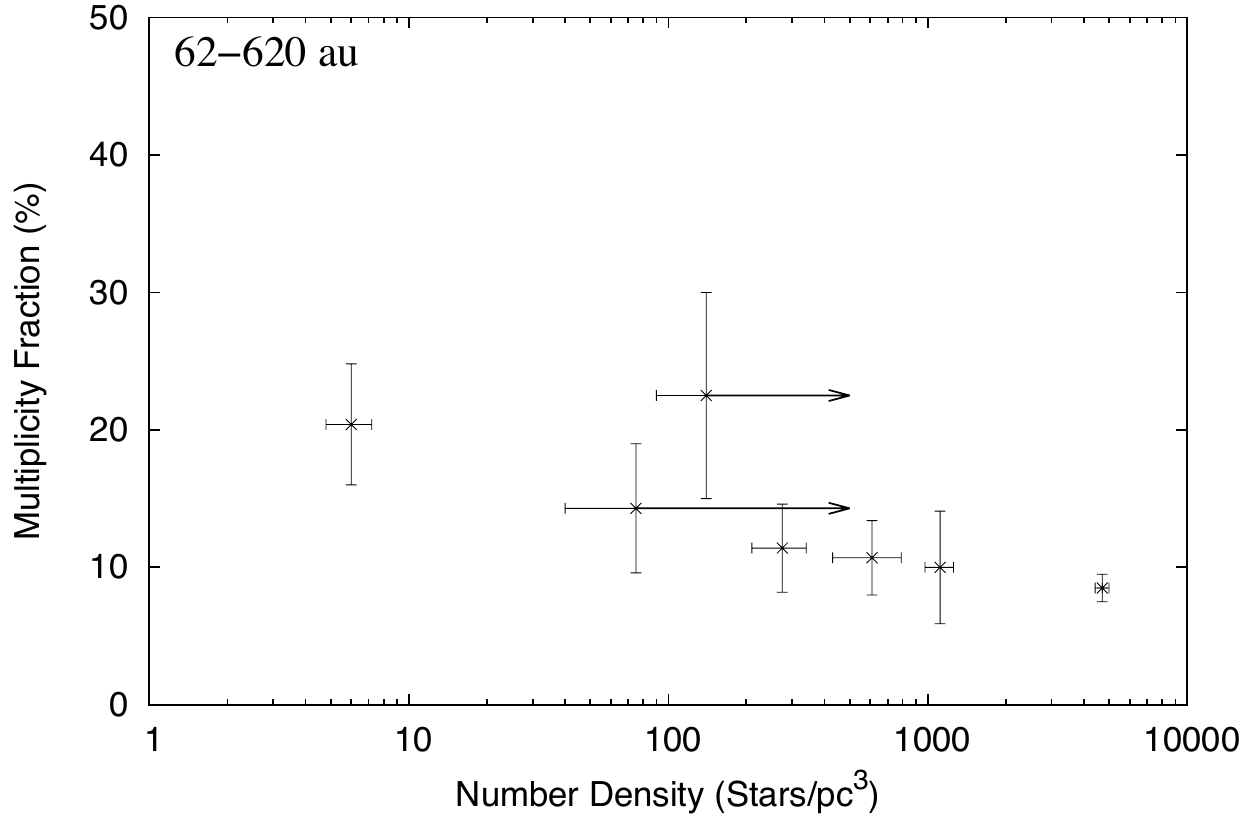}}
    \resizebox{\hsize}{!}{\includegraphics{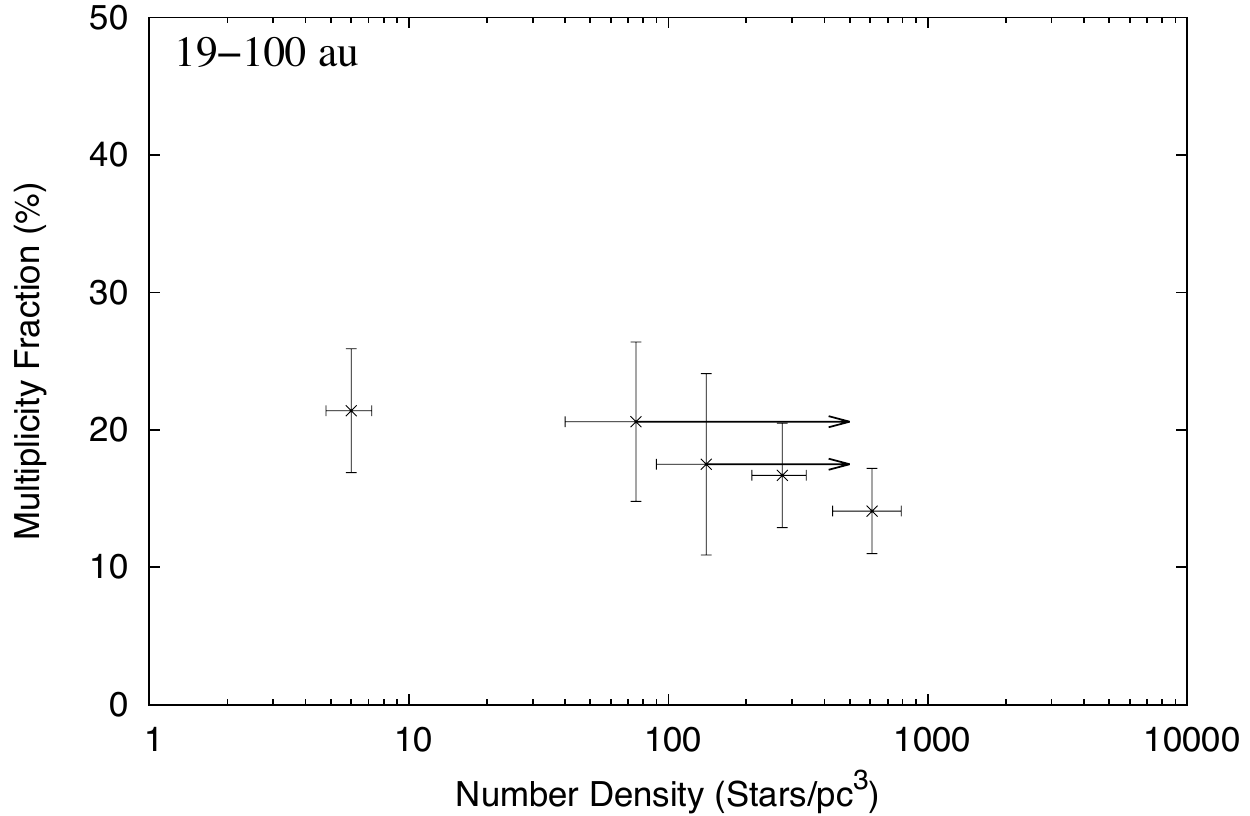}}
    \caption{The stellar multiplicity fractions of our regions as a function of 
    stellar density when we
    consider the same contrast cuts and stellar masses within the separation ranges 
    19--774\,au (top), 62--620\,au (middle), and 19--100\,au (bottom). 
    The
    densities are calculated within a projected distance of 0.25\,pc from the cluster centre, 
    except in the case of Taurus where a radius of 1\,pc is used. For CrA and USco these densities
    are lower limits due to the incomplete census of their stellar members.}
    \label{fig:binary_frac_vs_density}
\end{figure}

In Table\,\ref{tab:binfrac} and Fig.\,\ref{fig:binary_frac_vs_density} we present the multiplicity
fraction (MF) in each of our young regions and also give an estimate of the average volume densities
(in stars pc$^{-3}$).  The two new regions introduced in this paper are USco with a density of
$\gtrsim 80$ stars pc$^{-3}$, and CrA with a density of $>150$ stars pc$^{-3}$.  Due to the
incomplete memberships of both regions, these estimates are necessarily lower limits, as indicated
with arrows in Fig\,\ref{fig:binary_frac_vs_density}. CrA is a particularly problematic region in
that its membership is very poorly known and, with the relatively high levels of extinction, the
density may be significantly under-estimated. Although its density is extremely uncertain, we assume
that the multiplicity fraction and separation distribution are not biased by only sampling a subset
of the members.

\begin{table}
  \centering
  \caption{The densities (second column) and multiplicity fractions
    in each separation range (third to fifth column) for each of our
    young regions (named in the first column). The error on the
    multiplicity fraction is the poisson error, and the 
  number in brackets are the observed number of companions in each range
  for each region.}
\label{tab:binfrac}
\begin{tabular}{lcccc}
  \hline
  \hline
  Region 	& Density &  \multicolumn{3}{c}{Multiplicity fraction (per cent)}  	\\
   & (stars pc$^{-3}$)  &  19-100\,au & 19-774\,au & 62-620\,au  	\\
  \hline
  Taurus 	& 6               & $21\pm5$ (22)    & $37\pm6$ (38)  & $21\pm5$ (22)  \\
  USco 		& $\gtrsim 80$    & $21\pm6$ (13)    & $29\pm7$ (18)  & $14\pm5$ (9)   \\
  CrA 		& $> 150$         & $18\pm7$ (7)     & $38\pm10$ (15) & $23\pm8$ (9)   \\
  Cha~I	   & 280             & $16\pm4$ (19)    & $28\pm5$ (32)  & $11\pm3$ (13)  \\
  Oph       & 610             & $14\pm3$ (21)    & $24\pm4$ (36)  & $11\pm3$ (16)  \\
  IC\,348 	& 1100            & ... 	          & ...            & $9.7\pm4$ (6)   \\
  ONC 		& 4700            & ... 	          & ...            & $8.6\pm1$ (76)  \\
  \hline	
\end{tabular}
\end{table}

From our discussion of binary processing (see Sect.\,\ref{processing}), it might well be expected
that the multiplicity fraction will decrease with increasing density as encounters are closer, more
energetic, and more frequent. In Paper I, we found that the multiplicity fraction does not seem to
decrease significantly with density, rather that Taurus could well be an outlier with an unusually
high multiplicity fraction. With the addition of two more regions, we can revisit this analysis.

Consistent with the results of Paper~I, there is no {\em significant} correlation of multiplicity
fraction (MF) with density.  In Fig.\,\ref{fig:binary_frac_vs_density}, we show the MF against
density over three separation ranges. From the middle panel which shows the 62--620\,au range common
to all 7 regions, one might argue for a trend of decreasing multiplicity fraction with density with
a halving of the MF over three orders of magnitude in density between Taurus ($\sim 6$ stars
pc$^{-3}$) and the ONC ($\sim 5000$ stars pc$^{-3}$).  However, it should be noted that only Taurus
and CrA have an MF more than 1-sigma above that of the ONC, and the uncertain densities of CrA and
USco significantly impact the interpretation of Fig.\,\ref{fig:binary_frac_vs_density}.

While it could be argued that there is some trend of decreasing MF with density in the 19--774\,au and
62--620\,au separation ranges, in-line with the predictions of dynamical processing, it is very interesting
to note that the 19--100\,au separation range (bottom panel of Fig.\,\ref{fig:binary_frac_vs_density}) may
also show the same trend -- and this cannot be due to dynamical processing.  Unfortunately, we do not have
separations in this range for our two densest regions (USco and the ONC).

Rather frustratingly, the data on MFs with density give no clear results.  There may or may not be a
decrease in MF with density in the 62--620\,au range (which universal star formation would predict), 
and there may or may not be a decrease in MF with density in the 19--100\,au range (which universal star
formation would not predict).

\subsection{Comparisons of separation distributions}

The separation distributions of the regions provide another observational test of dynamical
processing.  To summarise: universal star formation would predict fewer wide binaries in denser
regions, and would predict identical separation distributions for close binaries.  It would also
predict that the sum of all young regions would be the same as the field.

\begin{figure}
    \resizebox{\hsize}{!}{\includegraphics{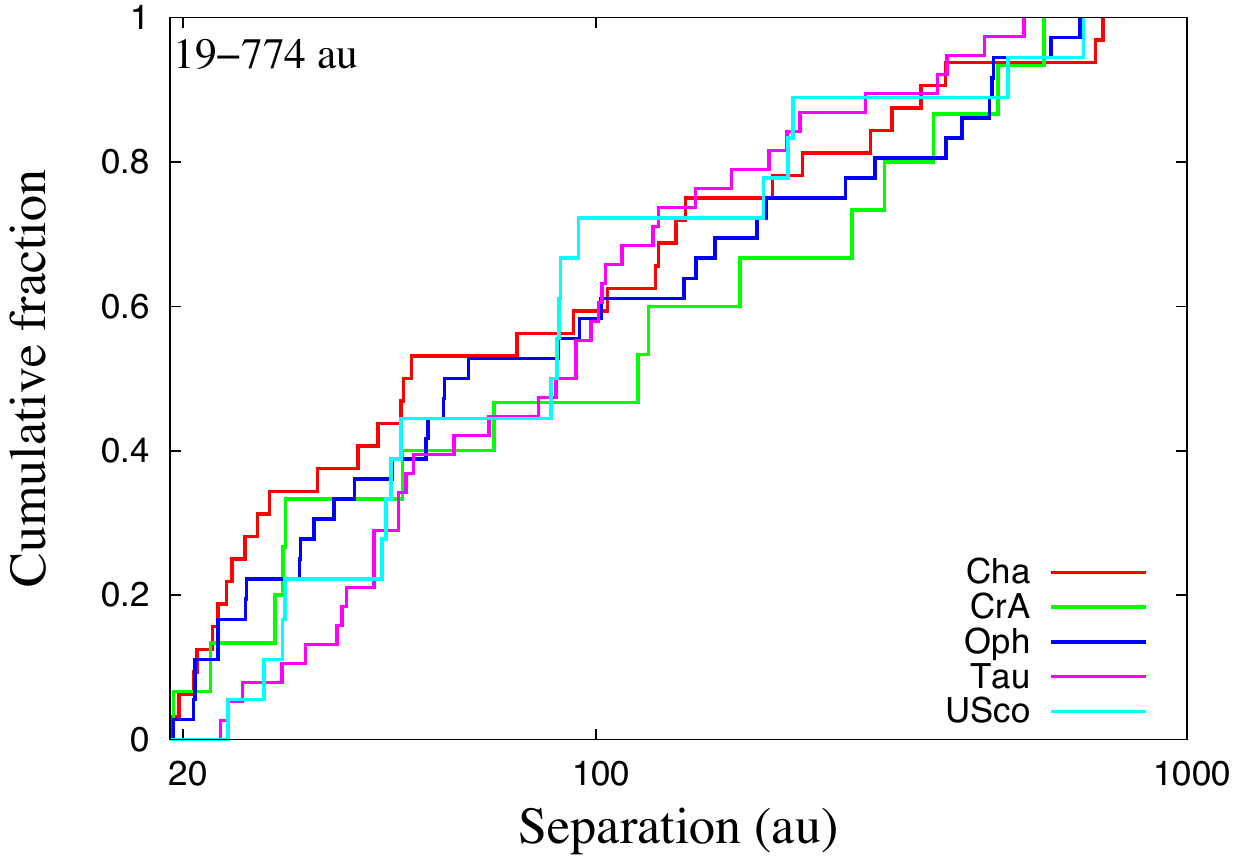}}
    \resizebox{\hsize}{!}{\includegraphics{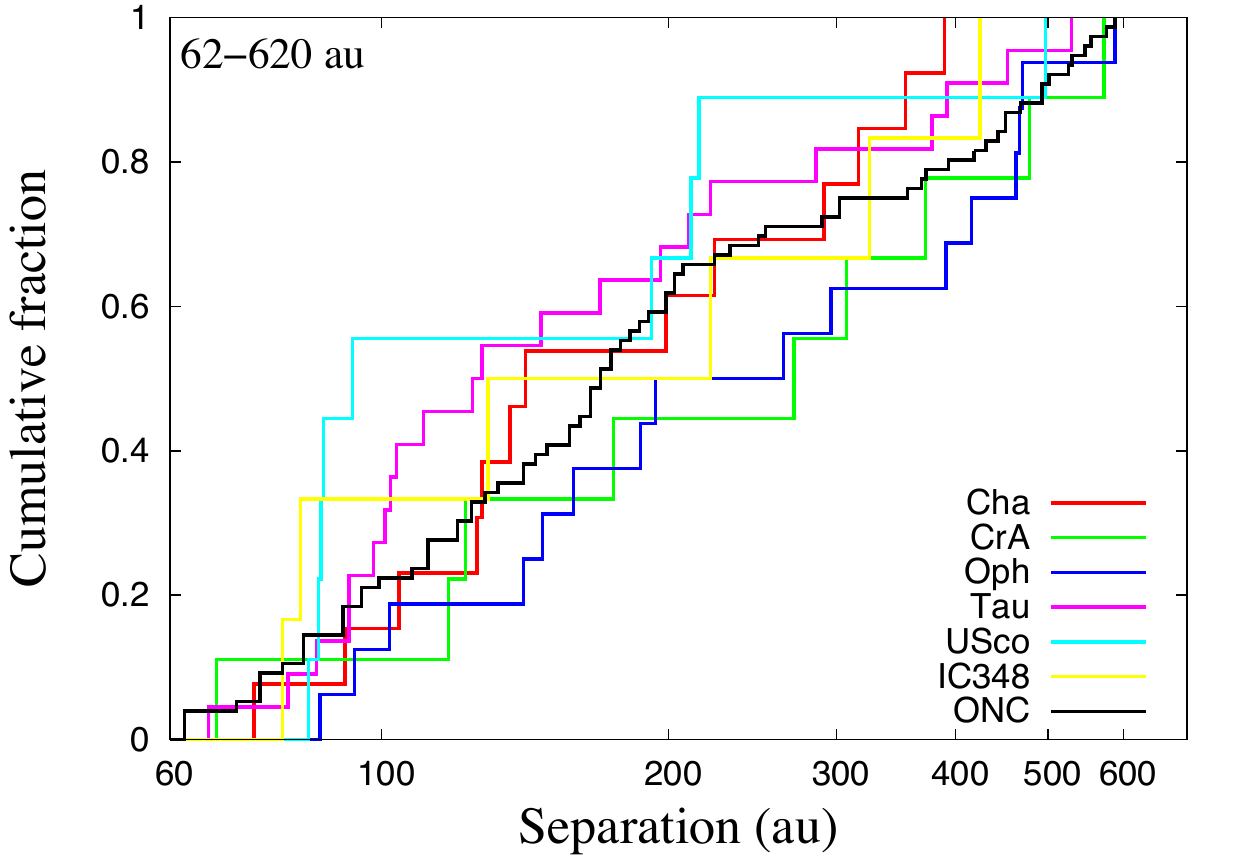}}
    \resizebox{\hsize}{!}{\includegraphics{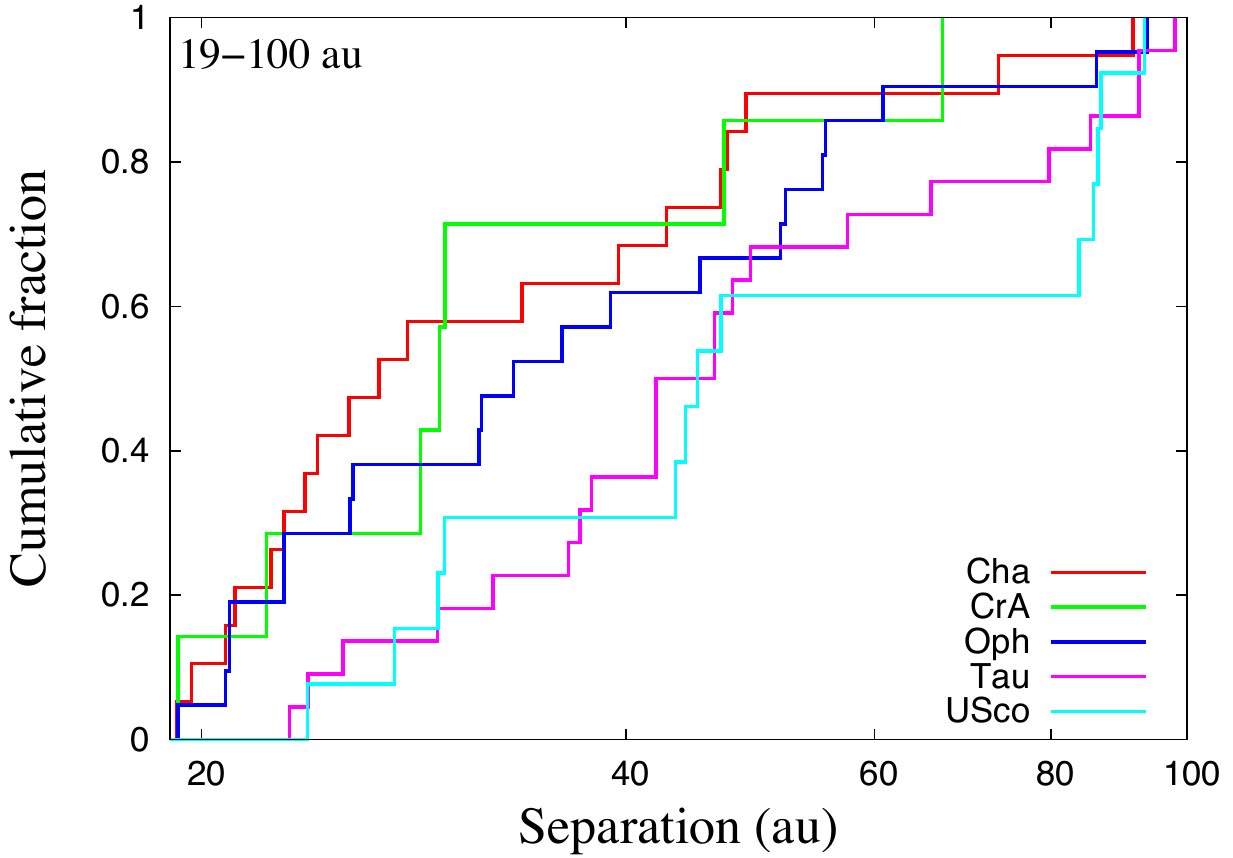}}
    \caption{The cumulative separation distributions for the companions 
    in our comparable samples for clusters within the separation ranges 
    19--774\,au (top), 62--620\,au (middle), and 19--100\,au (bottom).  
    The line colour 
    corresponding to each cluster is shown in the figures.}
    \label{fig:sep_distribution}
\end{figure}


\subsubsection{Comparisons between regions}

In Fig.\,\ref{fig:sep_distribution}, we compare the CDFs of separations for the young
regions. The top panel covers the widest range 19--774\,au for Cha, CrA, Oph, Tau, and
USco.  The middle panel shows the range 62--620\,au for all regions (Cha, CrA, Oph, Tau,
USco, IC~348, and the ONC) and the bottom panel is for the `pristine' range of 19--100\,au
for the same five regions as the 19--774\,au range. The most striking feature of these
figures, especially in light of our previous discussion of binary processing, is that all
of the separation distributions look remarkably similar, {\em except} in the limited
pristine range.

A KS test on the separation CDF of all pairs of clusters in both full separation ranges
(19--774\,au and 62--620\,au, top two plots in Fig.\,\ref{fig:sep_distribution}) shows no
statistically significant differences.  In the wider 19--774\,au separation range, the
lowest value of the KS probability is 0.15 (between Tau and Cha) which on its own is a very
marginal result, and given that we have 10 comparisons in total we would expect one to be
different at the 10\,per cent level.  In the more limited 62--620\,au range the lowest KS
probability is 0.08 (between Oph and USco), which again is at the level we would expect
from chance as we have 21 pairs to compare. Typically, the KS probability is between 0.2
and nearly unity. In many cases, low-number statistics makes finding a statistically significant difference
very difficult, but an examination by eye of Fig.\,\ref{fig:sep_distribution} also shows
that they all appear very similar.

Another interesting feature is that they all appear to be roughly flat in log-space
\citep[see][]{Kraus:2008,Kraus:2011}, although there is a tendency to have more low-separation 
(19--50\,au) binaries than expected from a completely log-flat distribution.

Very interestingly, in the 19--100\,au range of hard binaries there is tentative evidence for
differences between the separation distributions. The statistics in this range are limited,
but KS tests comparing regions to each other do suggest that Tau is different to both Cha (with
a KS probability of 0.025), and CrA (with a KS probability of 0.063).

Examination of the bottom panel of Fig.\,\ref{fig:sep_distribution} shows that both Tau and USco
appear to have a fairly flat separation distribution whilst Cha and CrA, in particular, have half of
their binaries between 19 and 30\,au.  However, these apparent differences should be treated with
some caution.

To investigate systems that should all be hard binaires, we have selected an arbitrary range of
separations, limited at 19\,au by observations and the choice of 100\,au as a reasonable hard-soft
boundary (see equation 2). Examination of the top panel of Fig.\,\ref{fig:sep_distribution}, which
shows the widest range available from the observations, shows that USco has a `jump' at just below
100\,au separations. Clearly it is possible to select various arbitrary small ranges within the
larger range that can find differences between separation distributions.  This obviously raises the
question as to what biases have been introduced by our (observationally constrained) choice of
separation ranges.

\subsubsection{Field binary separation distributions}

Separation distributions and multiplicity fractions are very often compared to the field.  The field
provides a useful reference distribution, but most importantly, the field must be the sum of all
past star formation in all different environments.

\citet{Duquennoy:1991} compiled an extensive sample for nearby field G-dwarfs.  They found that they
could fit the semi-major axis distribution (after correcting for observational effects) with a
log-normal distribution with mean $\mu_{{\rm log}_{10} a} \sim 1.5$ (roughly 30\,au), and variance
$\sigma_{{\rm log}_{10} a} \sim 1.5$.  A more recent study by \citet{Raghavan:2010} finds a similar
distribution.  These samples have a huge advantage of being volume-limited and complete for almost
all separations, therefore providing an ideal reference sample for primaries of $\sim 0.8$-$1.6
M_\odot$.

The separation distribution for stars other than G-dwarfs, however, is less certain.
\citet{Mayor:1992} found that K-dwarfs appear to have a similar separation distribution to G-dwarfs,
but with lower-number statistics.  \citet{Fischer:1992} compiled data for M-dwarfs from a variety of
sources: their results are possibly consistent with the G-dwarf separation distribution, but could
well favour a lower mean \citep[see also][]{Bergfors:2010}.  For A-stars it is possible that the
typical separation is wider than for G-dwarfs, but limited observed separation ranges and low-number
statistics mean that this is very difficult to confirm \citep{Kouwenhoven:2006}.

\subsubsection{Comparisons between the young regions and the field}

\begin{figure}
    
    \resizebox{\hsize}{!}{\includegraphics{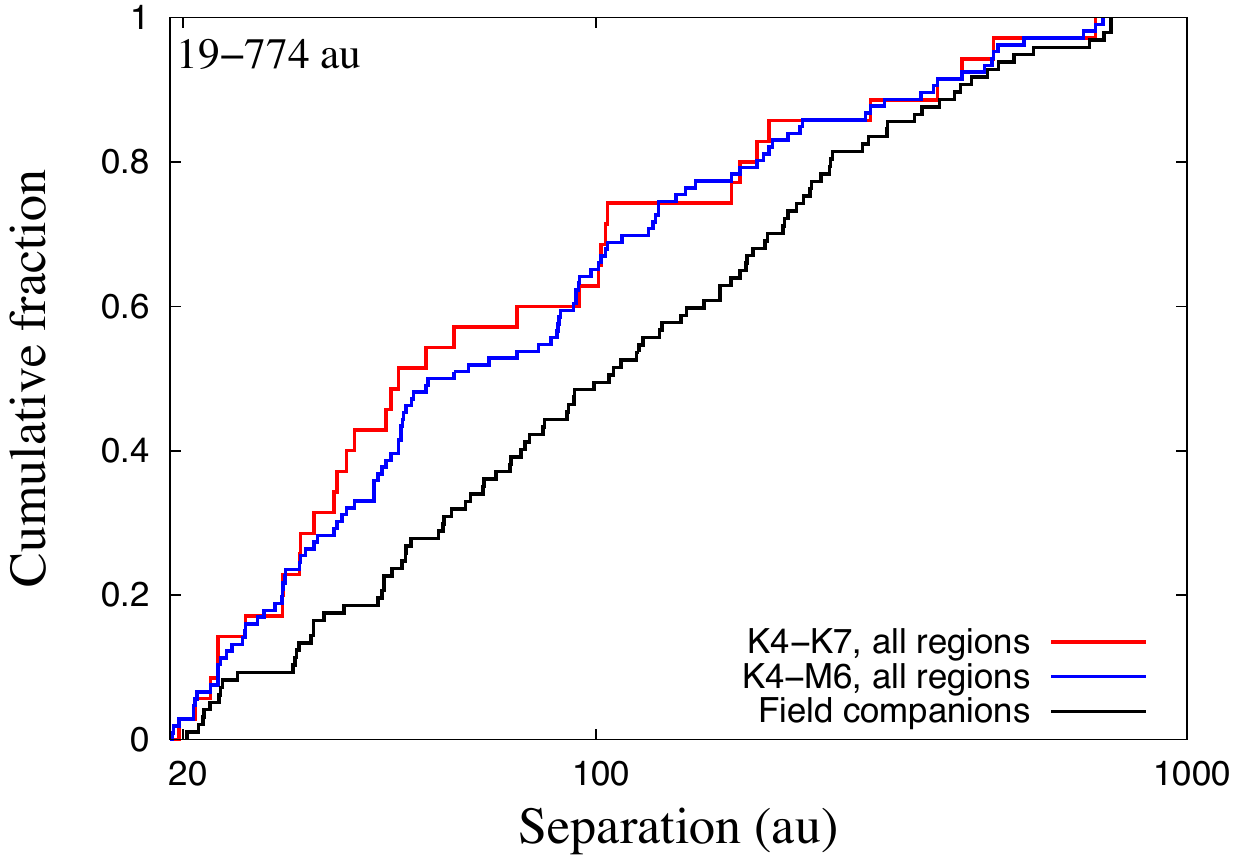}}
    \resizebox{\hsize}{!}{\includegraphics{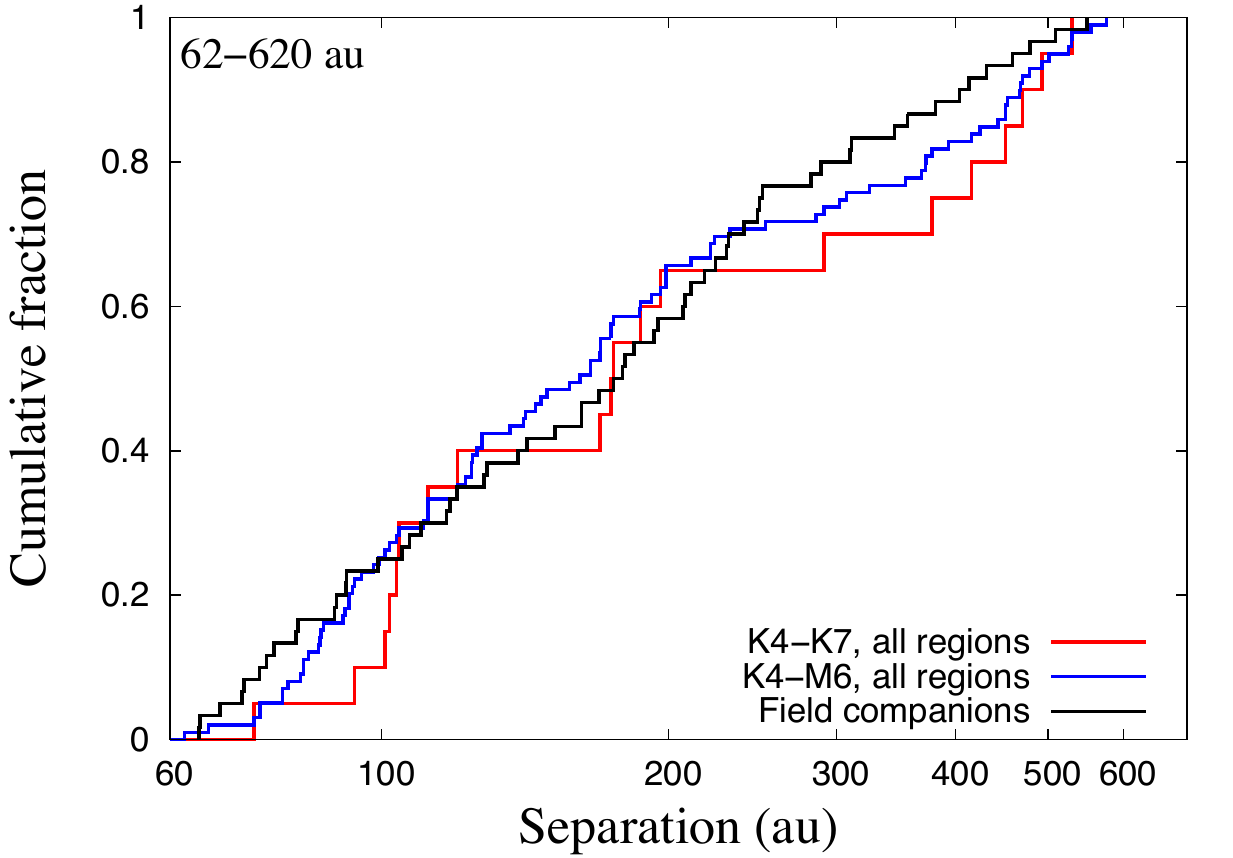}}
\resizebox{\hsize}{!}{\includegraphics{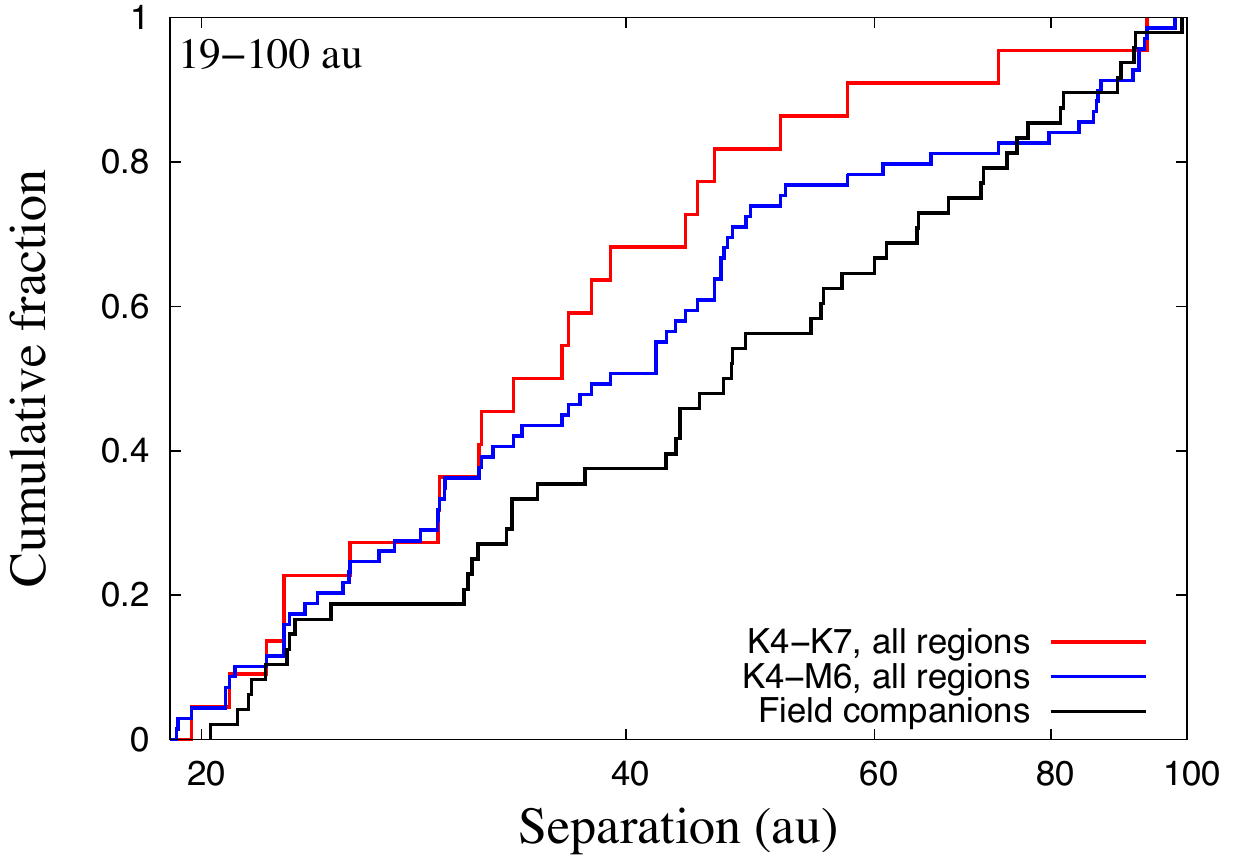}}
    \caption{The cumulative separation distributions for young binaries within the separation ranges
    19-774au (top), 62-620au (middle), and 19-100au (bottom), compared to the \citet{Raghavan:2010}
    observations of field binaries (black lines). The blue lines show the sample with K4--M6
    primaries. The red lines show the K4--K7 primaries which correspond to the F6--K3 range of the
    field primaries, i.e, the fairest comparison to the \citeauthor{Raghavan:2010} field survey.}
    \label{fig:field_distrib1}
\end{figure}

We have chosen to compare our separation distribution with that of \citet{Raghavan:2010}.  The primary stars
in the \citeauthor{Raghavan:2010} sample have spectral types from F6 to K3.  To better match the young region
sample we remove from their sample of 259 companions, 10 L and T-type brown dwarfs (to better-match our
sample of stellar companions) leaving 186 companions with measured separations, 48 with
spectroscopically-determined orbital elements and 15 with only estimates of the orbital period. For all of
the \citeauthor{Raghavan:2010} spectroscopic binaries, the periods are short enough that the instantaneous
separation will always be smaller than the lower limit to our separation ranges (19\,au) and for 13 of the 15
systems with estimated periods, those periods (or direct imaging) also imply maximum separations below
19\,au. We neglect the two systems for which we have no useful information (HD16673\,Aa,Ab and
HD197214\,Aa,Ab) and use the 186 companions with measured separations to make a comparison with our cluster
samples. Applying the same separation cuts to the Raghavan sample, we are left with 48, 60 and 97 companions
in separation ranges 19-100\,au 62-620\,au and 19-774\,au, respectively.

We add together all of the systems in our young regions to produce a single combined young sample. The
apparent similarities between all young regions in the 19--774\,au, and 62--620\,au ranges (see
above)  make this (we hope) not a completely unreasonable thing to do.

As noted in Sect.\,\ref{sect:Spectral type distributions}, the sample of young regions contains
primary stars with spectral types from G5 to M6, which will evolve to become main-sequence stars
with spectral types of $\sim$A0--M6. Although the majority of our sample is comprised of late-type
stars, the typically wider separations and higher multiplicity of high-mass stars
\citep[see][]{Kouwenhoven:2006} may bias any comparison to a field survey of solar-type stars.

We therefore make two comparisons between the young binaries and the \citet{Raghavan:2010}
sample of field binaries by selecting subsets of similar primary mass. The first comparison
involves young stars with spectral types K4--K7, which correspond to spectral types of
$\sim$F6--K3 at field ages (estimated from evolutionary models, see
Sec.\,\ref{sect:Spectral type distributions}), the closest possible match to the
\citeauthor{Raghavan:2010} targets. The second comparison includes all stars with spectral
types later than K4, \idest, all stars lower mass than $\sim$1.6$\Msun$. Although the
K4--M6 sample includes stars of lower-mass than the \citeauthor{Raghavan:2010} study, the
larger sample allows for a more statistically significantly comparison, and is interesting
to search for differences (or lack of) between the two young samples.

In Fig.\,\ref{fig:field_distrib1}, we show the comparisons between the separation distributions of
the complete low-mass sample (K4--M6, blue line), comparable primary mass sample (K4--K7, red line),
and field sample (black line).  This is done for the widest 19--774\,au sample (top panel), the
62--620\,au range encompassing all young regions (middle panel), and the 19--100\,au pristine range
(bottom panel).

\begin{figure}
    \resizebox{\hsize}{!}{\includegraphics{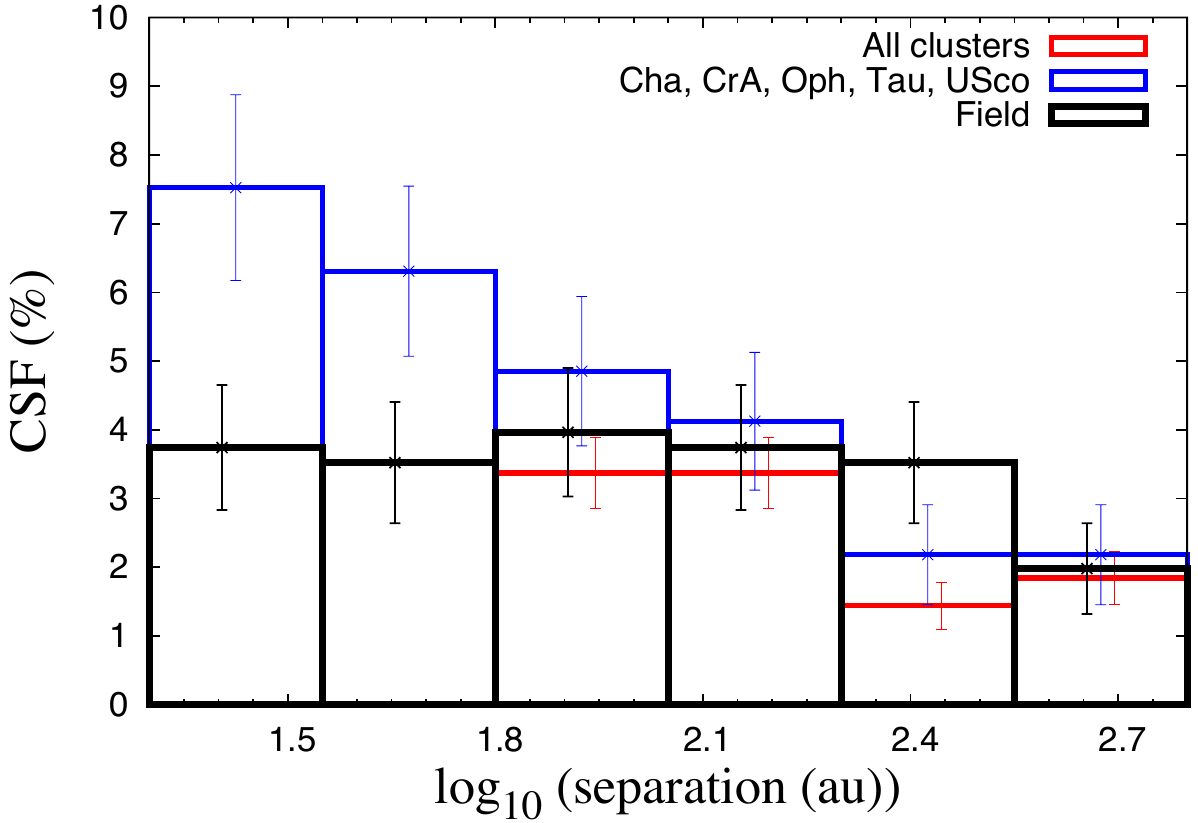}}
    \caption{The separation distributions against the {\em companion star fraction} (CSF) for all
    companions to K4--M6 primaries in regions where we have sensitivity to separations of
    19-774\,au (blue histogram) and 62-620\,au (red histogram) compared to the
    \citet{Raghavan:2010} field survey (black histogram). The CSF is used here for both the field and
    young regions to avoid having to choose which separation to plot for systems with 2 or more 
    companions.}
    \label{fig:bfsepbins}
\end{figure}

To supplement Fig.\,\ref{fig:field_distrib1}, the companion star fractions (CSFs) of the young
samples in bins of separations are plotted in Fig.\,\ref{fig:bfsepbins}, where the black histogram
shows the field CSF, the red histogram shows CSFs for all clusters in the limited 62--620\,au range
(cf. bottom panel of Fig. \,\ref{fig:field_distrib1}), and the blue histogram shows the CSFs for the
five regions with uniform coverage in the 19--774\,au range.

There are a number of very interesting features of Figs.\,\ref{fig:field_distrib1}
and \ref{fig:bfsepbins}.  Firstly, the two young samples (K4--M6, blue and K4--K7, red) in
Fig.\,\ref{fig:field_distrib1} are similar in all cases.  The subset restricted to higher-mass
primaries has fewer stars within it as most of the primaries in the young regions are relatively
low-mass, but a KS test does not distinguish between the young region samples in any of the three
ranges.

Secondly, the field separation distribution is significantly different to the young samples in two of
the three ranges shown in Fig.\,\ref{fig:field_distrib1}. In the 19--100\,au range the KS probability
comparing the field and the K4--M6 sample is 0.16, whilst for comparable primary masses (K4--K7
sample) it is only 0.05. In the 19--774\,au range the KS probability comparing the field and the
K4--M6 sample is 0.01, whilst for comparable primary masses it is 0.03. In the 62--620\,au range,
however, the KS probability comparing the field and the K4--M6 sample is 0.65, whilst for comparable
primary masses it is 0.81.

This illustrates a potential danger when using limited ranges of separation with a cumulative
distribution plot.  An interpretation using only the 62--620\,au range in the middle panel of
Fig.\,\ref{fig:field_distrib1}, would lead to the conclusion that the field and young cluster
separation distributions were indistinguishable, but using the 19--774\,au range in the top panel
suggests very significant differences.

Examination of the top panel of Fig.\,\ref{fig:field_distrib1} shows the source of the apparent
discrepancy.  Around 50\,per cent of young binaries in the range 19--774\,au have a companion closer
than $\sim$50\,au, whilst only around 25\,per cent of field stars do.

In Fig.\,\ref{fig:bfsepbins}, we can compare restricted separation ranges by binning the data.  In
this figure, it is clear that the over-abundance of binaries with separations $< 60$\,au in the
young regions compared to the field is real.  In the first two bins of Fig.\,\ref{fig:bfsepbins}, we
see that the young regions have roughly twice the binary fraction found in the field.  However, it
also illustrates that these bins only contain data from the lower-density regions -- IC~348 and the
ONC have no observations at these close separations.

\section{Discussion}

To address the question of the universality of star formation, we examine if multiplicity fractions
and separation distributions (within given separation ranges) vary from region to region, and
whether they differ from the field.

We have presented a sample of companions in 7 young regions in which we have, as far as possible,
produced comparable samples with the same selection criteria.  This allows a comparison of
multiplicity fractions and separations in two ranges: 19--774\,au for 5 clusters (Cha, CrA, Oph,
Tau, and USco) and 62--620\,au for all clusters (also including IC~348, and the ONC).  In addition,
we selected a limited 19--100\,au range in which all of the systems are expected to be unaffected by
dynamical processing.

There is a possible weak trend of decreasing multiplicity fraction with increasing density in all
separation ranges, including, unexpectedly, the 19--100\,au unprocessed range. In the 62--620\,au
range, the separation distributions of all our young regions are similar to each other and to the
field. In the wider 19--774\,au range the separation distributions of the five young regions are
similar to each other, but significantly different to the field due to an excess of close binaries
($\lesssim$50\,au, roughly twice what is found in the field). The only range in which there is any
statistical evidence for variations between separation distributions of individual young clusters is
in the 19--100\,au range.

We reiterate here that we are considering a limited range of binary separations of 19--774\,au in
which we can compare different regions.  In particular, we only have data for all regions in a very
limited range of 62--620\,au \citep[set by the observational constraints of the ONC sample
of][]{Reipurth:2007}.

\subsection{Non-universal binary formation?}

The `standard model' of binary formation and processing suggests that the multiplicity fractions and
separation distributions at birth are the same in all regions
\citep[e.g.][]{Kroupa:1995b,Kroupa:1995a,Kroupa:1995c,Goodwin:2005,Parker:2009, Marks:2011a}.  It
suggests that any differences between regions are due to dynamical processing which is more
effective in denser regions, reducing both the multiplicity fraction and preferentially destroying
wider binaries (see Sect.\,\ref{processing}).  Binary destruction is a rather stochastic process in
the intermediate regime and so the resulting separation distributions and multiplicity fractions may
vary somewhat, but the general picture of lowered multiplicity fraction and preferential destruction
of wider binaries in denser environments usually holds \citep{Parker:2012}. Based on four main
observations, \emph{our analysis of the observational data strongly suggests that this standard
model is incorrect}.

Firstly, a trend of decreasing multiplicity fraction with density is expected in the standard model.
However, we find that if any trend exists (for which the evidence is weak), it could also be present
in the closest 19--100\,au binaries where it cannot be due to dynamical processing (see
Sect.\,\ref{processing}).

Secondly, denser regions should process their wider binaries more strongly than low-density regions.
It would be expected that both IC~348 and the ONC (the densest regions) should show a separation
distribution with fewer wider ($>$300\,au) binaries than low-density regions.  Instead, all regions
are indistinguishable in the 62--620\,au range.  It is possible, but unlikely, that {\em both} IC~348
and the ONC are unusual in the way they have processed their binaries -- see \citet{Parker:2012}.

Thirdly, the separation distributions of close binaries in the 19--100\,au range cannot have been
dynamically altered and consequently should always look the same (in different regions and in the
field) in the standard model.  However, this is the one region in which there {\em is} evidence for
statistically significant differences.

Finally, the field is expected to match the sum of all star-forming regions.  In particular,
non-processable binaries in the 19--100\,au range should match the field, but
Figs.\,\ref{fig:field_distrib1} and \ref{fig:bfsepbins} show that they do not.  We note that we do
not have data for our densest regions in the 19--100\,au range, but for the sum of our regions to
match the field these two regions would have to have a very significant {\em lack} of 19--100\,au
binaries to balance the overabundance in our low-density regions.

This set of observational data strongly suggests that {\em binary formation is not the same
everywhere} and therefore {\em star formation is not a single universal process}.

Probably the strongest evidence for this is the over-abundance of 19--100\,au binaries in the
low-density regions compared to the field. This is extremely problematic as such binaries cannot be
destroyed by dynamical processing in any nearby environment. Therefore to match the field
population, some regions {\em must} under-produce close binaries, or the nearby young regions
on which we base our interpretation are not the source of most stars in the field.  This second
interpretation would be rather worrying as we base much of our understanding of star formation on
these regions and to discover they are not the source of most field stars would be problematic (to
say the least).

Another unexpected observation is that the dense regions have the same distribution of intermediate
binaries as the low-density regions.  They have marginally fewer binaries in the 62--620\,au range
than the lower-density regions, but the same fraction of their binaries are in the 62--300\,au range
as in the 300--620\,au range.  This is unexpected as denser regions should be much more effective at
processing 300--620\,au binaries than low-density regions. If anything, this suggests that
high-density regions must produce {\em more} intermediate binaries than low-density environments
which are then processed to a similar distribution as the low-density environments.

Recent observations suggest that this trend is not solely present in our compilation of data. 
\citet{Biller:2011} find an excess of 10--50\,au brown dwarf and VLMS binaries in USco compared to
the field.  \citet{Kraus:2011} also find an excess of 5--100\,au binaries in Taurus compared to the
field for low-mass (0.25--0.7$M_\odot$) stars, but not for higher-mass stars.

A possible model to explain these observations is that the denser an environment is, the wider the
binaries it produces.  That is, low-density regions over-produce close binaries compared to the
field, but high-density regions under-produce them (and the sum results in the field values as no
processing can occur).  The similarities between separation distributions and multiplicity fractions
at 300--620\,au between high- and low-density regions might suggest that high-density regions
over-produce such binaries compared to the field and low-density regions.  This is because such
binaries are destroyed more effectively in high-density regions, therefore to look the same now,
they must have started differently.

\subsection{Altering close binaries}

If the field is the sum of star-forming regions, then the over-abundance of close (and unprocessed)
binaries in nearby low-density star-forming regions compared to the field must be explained. In
terms of dynamical destruction through encounters, so-called `super star clusters', such as the
Galactic Arches, Trumpler 14 and Westerlund 1, or R136 in the LMC, may be dense enough to affect
binaries with separations $<$100\,au.  If these regions do process such close systems, and they all
evaporate into the field, then the under-abundance of close binaries in the field compared to
low-mass star-forming regions could be reconciled.

We have data on $<100$\,au binaries only for regions with total masses of a few $\times 10^2$
M$_\odot$.  Naked clusters appear to have a cluster mass function of the form $N(M_{\rm cl}) \propto
M_{\rm cl} ^{-2}$ \citep{Lada:2003}.  If this mass function extends to embedded clusters, it suggests
that an equal mass of stars forms in each decade of cluster mass.  Therefore an equal mass of stars
forms in super star clusters ($\sim 10^5$ M$_\odot$) as in our low-mass regions ($\sim 10^2$
M$_\odot$).  At first sight this may appear to solve the over-abundance of close binaries, however in
a universal star formation model {\em all} stars must form with the low-density, unprocessable
over-abundance we see in the $\sim 10^2$ M$_\odot$ regions.  However, close binaries can only be
processed in very high density regions with masses $> 10^4$ M$_\odot$, thus requiring a bi-modal
density distribution in regions forming half of stars in low-density environments and half in very
high density environments.

Finally, we note a recently proposed mechanism for the destruction of close binaries.
\citet{Korntreff:2012} suggest that binaries form with a log-flat separation distribution. The wide
binaries are processed dynamically in the cluster environment, whereas the close binaries decay due
to dynamical friction with gas in the cluster, thereby sculpting the log-flat distribution into the
log-normal observed in the field. For the orbital decay mechanism to be effective, a high gas
density is required (10$^4$--10$^6$ cm$^{-3}$), which presumably may be present in very massive
clusters. However, we note that this mechanism is most effective at low separations ($<$10\,au), and
it is difficult to see how this could destroy or alter the required proportion of binary systems in our
sample, which have separations in the range 19--100\,au.


\section{Conclusions}

By collating and analysing binary statistics for seven nearby young regions we have created
comparable samples of companions to stars with similar masses, separations from 19 to 774\,au and
contrasts of $\Delta K \le 2.5$. We compare all seven regions with each other and with the field in
the 62--620\,au separation range.  For the wider 19--774\,au range we cannot include our densest
regions (IC~348 and the ONC), nor can we for a restricted 19--100\,au range in which dynamical
processing should be unimportant.

Our results can be summarised as follows.
\begin{itemize}

\item There is either a weak trend or no trend of decreasing multiplicity fraction with density in
all separation ranges, including the unprocessed 19--100\,au range.

\item The separation distributions of all regions are statistically indistinguishable from one
another, except in the 19--100\,au separation range.

\item The multiplicity fractions and separation distributions of the young regions are very
different to the field in all but the 62--620\,au separation range. Specifically, there is an excess
of close binaries ($<$100\,au) in these nearby regions compared to the field.

\end{itemize}

Our conclusion from these results is that binary formation is not universal and consequently {\em
the star formation process is not universal}.  The 19--100\,au range in the low-density regions has
not been, and will not be, dynamically processed, yet it is inconsistent with the separation
distribution and multiplicity of comparable field stars. Only the densest galactic clusters could
process some of their sub-100\,au binaries, but to explain the discrepancy half of all field
binaries must originate in very massive clusters.  This excess of close binaries in low-density
regions compared to the field must mean that other regions under-produce close binaries, or
that the regions we have analysed are atypical in some way.

To confirm these intriguing results, more observations, especially probing smaller separations, are
required. CrA may be a particularly fruitful target as it is a relatively nearby region
($\sim$130\,pc) with a Taurus-like multiplicity, but at significantly higher density. A more
complete membership census would allow a firmer determination of its density and could demonstrate
that denser regions do not necessarily have fewer multiple systems. However, it must be emphasised
that care must be taken in ensuring that comparisons apply the same selection effects (separation 
range, primary mass, and sensitivity to companions), in particular between different regions, but
also between regions and the field.

\section*{Acknowledgments} 

We acknowledge support for this work from European Commission Marie Curie Research Training Network
CONSTELLATION (MRTN-CT-2006-035890) and the Exeter Astrophysics visitor grant. RRK and JP are
supported by a Leverhulme research project grant (F/00 144/BJ).  JP is also supported by a Science
and Technology Facilities Council grant (ST/11002707/1). Part of this work was completed at the
International Space Science Institute in Bern Switzerland as part of an International Team.


\bibliographystyle{mn2e_var}
\setlength{\bibhang}{2.0em}
\setlength\labelwidth{0.0em}
\bibliography{bib}

\bsp
\label{lastpage}

\end{document}